\documentclass[twocolumn, final]{svjour3}          

\makeatletter
\expandafter\let\csname opt@amsmath.sty\endcsname\relax
\makeatother

\usepackage{multirow}
\usepackage{graphicx}
\usepackage{bm}
\usepackage{amsmath}

\begin{document}

\title{Time irreversibility in reversible shell models of turbulence\thanks{Version accepted for publication (postprint) on Eur. Phys. J. E (2018) 41: 48 -- Published online: 6 April 2018}\thanks{Contribution to the Topical Issue ``Fluids and Structures:
Multi-scale coupling and modeling'' edited by Luca Biferale,
Stefano Guido, Andrea Scagliarini, Federico Toschi.}}


\author{Massimo De Pietro         \and
        Luca Biferale \and Guido Boffetta \and Massimo Cencini 
}


\institute{Massimo De Pietro and Luca Biferale \at
              Dipartimento di Fisica and INFN, Universit\`a di Roma “Tor Vergata”, Via Ricerca Scientifica 1, 00133 Roma, Italy \\
           \and
           Guido Boffetta
            \at
Dipartimento di Fisica and INFN, Universit\`a di Torino, via P. Giuria 1, 10125 Torino, Italy
\and
Massimo Cencini \at
              Istituto  dei  Sistemi  Complessi,  CNR,  via  dei  Taurini  19,  00185  Rome,  Italy  and  INFN  “Tor  Vergata” \\
              Tel.: +39-06-49937453\\
              Fax: +39-06-493440\\
              \email{massimo.cencini@cnr.it} 
              }

\date{}

\maketitle

\begin{abstract}

Turbulent flows governed by the Navier-Stokes equations (NSE) generate
an out-of-equilibrium time irreversible energy cascade from large to
small scales. In the NSE, the energy transfer is due to the nonlinear
terms that  are formally symmetric under time reversal. As for the dissipative
term: first it explicitly breaks time reversibility; second it
produces a small-scale sink for the energy transfer that remains
effective even in the limit of vanishing viscosity.  As a result, it
is not clear how to disentangle the time irreversibility originating
from the non-equilibrium energy cascade from the explicit
time-reversal symmetry breaking due to the viscous term. To this aim,
in this paper we investigate the properties of the energy transfer in
turbulent Shell models by using a reversible viscous mechanism,
avoiding any explicit breaking of the $t \rightarrow -t$ symmetry. We
probe time-irreversibility by studying the statistics of Lagrangian
power, which is found to be asymmetric under time reversal also in the
time-reversible model. This suggests that the turbulent dynamics
converges to a strange attractor where time-reversibility is
spontaneously broken and whose properties are robust for what concerns
purely inertial degrees of freedoms, as verified by the anomalous
scaling behavior of the velocity structure functions.
 
\end{abstract}
\section{Introduction\label{intro}}
Incompressible fluid motion is governed by the Navier-Stokes equations
(NSE):
\begin{equation}
\partial_t \bm u+\bm u\cdot \bm \nabla \bm u=-\bm \nabla p + \nu \Delta \bm u
+\bm F \,
\label{eq:NS}
\end{equation}
where $\bm u(\bm x,t)$ is the velocity field, $p$ the scalar pressure
ensuring $\bm \nabla \cdot \bm u=0$, $\nu$ the viscosity and $\bm F$
represents an external stirring force.  In the absence of viscosity
($\nu=0$), the NSE are invariant under time reversal, i.e. the
simultaneous transformation $\bm u \to -\bm u$ and $t \to -t$,
provided $\bm F$ respects this symmetry.  This means that if at time
$t$ we reverse the fluid velocity, the flow will trace back its
evolution.
 
The effects of viscosity are particularly subtle for turbulent flows
at high 
Reynolds numbers:
\begin{equation}
Re = \frac{U_LL}{\nu}
\end{equation}
where $L$ is the characteristic length of the flow and $U_L$ the
associated velocity. Fully developed turbulence corresponds to the
fluid state realized in the limit $Re \to \infty$, which is equivalent
to $\nu \to 0$ for fixed large scale flow configuration. As a result,
one could naively think that in this limit the dynamics becomes
reversible with zero mean energy flux.  This is not observed: it is an
empirical fact that in three dimensions turbulence dissipates energy
at a finite average rate, $\langle \varepsilon\rangle$, independently
of the value of viscosity, a fact known as the {\it dissipative
  anomaly} \cite{frish_turbulence}.  Thus, viscous effects play a
singular role in the dynamics of turbulent flows. 
Moreover,
it is also known that the Euler equations ($\nu =0$) can develop weak
solutions \cite{de2010admissibility} that do not conserve energy as
already conjectured by Onsager in the 40's. As a consequence, at least
formally, there is no need of a viscous sink to absorb energy in three
dimensional fluids.  As a result, we still lack a fundamental
understanding of time irreversibility in the strongly
out-of-equilibrium energy cascade (from large to small scales)
observed in 3D turbulent flows.  In particular, it is not clear how to
disentangle the effects due to the explicit time reversal symmetry
breaking introduced by the viscous term from the breaking due to the
attractor selected by the non equilibrium dynamics, similarly to what
happens for macroscopic time irreversibility in the thermodynamical
limit of systems with a time reversible microscopic dynamics
\cite{rose1978fully,falkovich2006}.

In this paper we further investigate this fundamental issue by
studying the evolution of a family of dynamical models for the NSE
equipped with a fully time-reversible viscosity, elaborating an
original idea proposed by Gallavotti at the end of the '90s
\cite{gallavotti1996equivalence,Gallavotti1997Dynamical} (see also
\cite{gallavotti2004lyapunov,gallavotti2014equivalence}) and never
fully checked in strongly out-of-equilibrium systems as the turbulent
energy cascade. In a nutshell, the idea consists in allowing the
viscosity to change such as some 
global quantity is exactly conserved, for example by fixing the total
energy or enstrophy of the flow.
In this way, we move from the original dynamics where viscosity
is fixed and the total energy (or enstrophy) is chaotically changing
in time around some stationary value to a system where viscosity is
oscillating with fixed energy (enstrophy).  Loosely speaking, we are
playing a similar game when moving from canonical to microcanonical
ensembles in equilibrium statistical mechanics. Here the system will
be out-of-equilibrium, and it is far from trivial to prove the
equivalence of the two descriptions.

In the original NSE, time-reversal symmetry breaking can be easily
revealed by studying multipoint Eulerian or Lagrangian correlations, as
for the case of the third order moment of the velocity increments in
the configurational space \cite{frish_turbulence} or the relative
dispersion of two-or-more particles
\cite{sawford2005comparison,jucha2014time,biferale2005multiparticle}.
Remarkably, irreversibility manifests also in the dynamics of a single
fluid element as recently found in \cite{xu2014,xu2014b} (see also
\cite{cencini2017time}).  Fluid elements, or tracers, evolve according
to the dynamics $\dot{\bm x}=\bm v(t)\equiv \bm u(\bm x,t)$. By
inspecting experimental and numerical tracer trajectories it was
discovered that the Lagrangian kinetic energy,
$\mathcal{E}(t)=\frac{1}{2}v^2(t)$, is dominated by events in which it grows
slower than it decreases \cite{xu2014}. As a consequence, the rate
of the kinetic energy change (Lagrangian power),
\begin{equation}
p(t)\equiv\dot{\mathcal{E}}(t)=\bm
v(t)\cdot \bm a(t)
\label{eq:pnse}
\end{equation}
where $\bm a$ is the particle's acceleration, is characterized by a
skewed distribution with $\langle p^3\rangle$ negative and scaling
with a power of the Reynolds number. Such asymmetry is directly linked
to time irreversibility \cite{xu2014}.  These features have been found
also in compressible \cite{grafke2015} and two-dimensional turbulence
\cite{xu2014,piretto2016irreversibility}.  It should be emphasized
that the skewness of the Lagrangian power is also relevant to more applied
issues such as the stochastic modelisation of single particle
transport in turbulent environmental flows \cite{sawford}.

In \cite{cencini2017time}, the authors have investigated the Lagrangian power
statistics by means of direct numerical simulations (DNS) of the NSE
(\ref{eq:NS}) and of shell models of turbulence
\cite{biferale2003shell,bohr2005}.  By looking at observables that are
sensitive to the asymmetry of the probability distribution function
(pdf), we found that both the symmetric and the time asymmetric
components do scale in the same way in the DNS data and the scaling
properties can be rationalized within the framework of the
multifractal (MF) model of turbulence, which is blind to time-symmetry
\cite{FP1985,benzi1984multifractal}. {Because the measured asymmetry
  is very small and the Reynolds numbers naturally limited by the
  numerical resolutions in three dimensions, we studied in the same
  paper also shell models, where a clear difference in scaling among
  symmetric and anti-symmetric components was observed.  Not
  surprisingly, by applying the same multifractal theory valid for NSE
  it is possible to capture the symmetric part of the Lagrangian power
  statistics only. The latter result suggests that shell models are a
  good playground for asking precise questions concerning the relative
  importance of (time) symmetric \textit{vs} asymmetric components of
  the Lagrangian power pdf at Reynolds numbers otherwise not
  achievable in the NSE case.

In the following we extend the study of time irreversibility initiated
in \cite{cencini2017time} by using a family of
\textit{time-reversible} shell models, obtained by modifying the
viscosity according to Gallavotti's idea.  Besides the academic
interest on such kind of models, it is important to remark that
reversible dissipative terms have been also used in Large Eddy
Simulations (LES) of the NSE
\cite{she1993constrained,carati2001modelling,fang2012time,Jimenez2015}.
Therefore, investigating such reversible equations, even in the
simplified framework of shell models is of interest for the more
general issue of developing effective models for the small scales of
turbulence (see, e.g., the discussion in \cite{fang2012time}).

Comparing \textit{vis a vis} the dynamics of the irreversible shell
model (ISM) with its reversible (RSM) variant offers us a unique
possibility to deepen the understanding of the Lagrangian power
statistics, and its connection with irreversibility. In particular, we
show here that for RSM, the time reversibility is
\textit{spontaneously} broken due to the non-equilibrium character of
the dynamics. We also show that RSM share the same statistical
properties of ISM for all inertial degrees-of-freedoms, those that are
not directly affected by the properties of the specific
time-reversible viscous mechanisms, while dissipative statistics is
different. Our results suggest that time-irreversibility is a robust
property of the turbulent energy transfer, and that it is
spontaneously broken on the attractor selected by the dynamics.

The paper is organized as follows. In Section \ref{sec:shellmodels} we
briefly recall the idea behind shell models, describe the particular
model considered and introduce its reversible formulation. We end the
section recalling how Lagrangian statistics can be studied within the
shell model framework. In Section \ref{sec:confronto} we compare the
statistics of RSM and ISM, in particular we focus on the structure
functions and their scaling behavior in the inertial range. We end the
section discussing the small-scale properties of the RSM, where the
modified dissipation acts more strongly.  Section \ref{sec:lagpow} is
devoted to the Lagrangian power statistics. We first briefly summarize
previous findings and then focus on the results of simulations of the
two models. Section \ref{sec:conclusions} is devoted to
conclusions. In Appendix \ref{sec:app_a} we provide some details on
the numerical simulations of the RSM, while in Appendix~\ref{app:MF}
we summarize the basics of the multifractal model for turbulence and
its application to Lagrangian statistics.

\section{Irreversible and reversible shell models \label{sec:shellmodels}}
Shell models are finite dimensional, chaotic dynamical systems
providing a simplified laboratory for fundamental studies of fully
developed turbulence
\cite{frish_turbulence,bohr2005,biferale2003shell,ditlevsen2010turbulence}.
These models have been introduced as drastic simplifications of the
NSE and, remarkably, found to share with them many non-trivial
properties encompassing the energy cascade, dissipative anomaly, and
intermittency with anomalous scaling for the velocity statistics.  In
this section we describe the so called ``Sabra'' shell model
\cite{Lvov_1998_improved_shellmodels} and introduce a variant of it
where the dissipative term is modified as proposed in
\cite{gallavotti1996equivalence,Gallavotti1997Dynamical} in order to
obtain formally time reversible equations. We end the section by
showing how the shell model can be used to study Lagrangian power
statistics.

\subsection{Standard (irreversible) Sabra shell model (ISM) \label{sec:sabra}}
The Sabra shell model \cite{Lvov_1998_improved_shellmodels} is a
modified version of the well known Gledzer-Ohkitani-Yamada model
\cite{gledzer1973system,ohkitani1989temporal} for which anomalous
scaling was first observed \cite{jensen1991intermittency}.  As typical
for shell models, the dynamics is defined over a discrete number of
shells in Fourier space arranged in a geometric progression $k_n=k_0
\lambda^{n-1}$ with $n=1,\ldots,N$ (with $k_0=1$ and $\lambda=2$ in
our simulations). A complex velocity variable $u_n(t)$ is
considered for each shell, which can be interpreted as the velocity
fluctuation (eddy) at scale $k^{-1}_n$.  The Sabra model equation for
$u_n$ reads:
\begin{equation}
\begin{aligned}
\dot{u}_n =& -\nu k_n^2 u_n +ik_n (a\lambda u_{n+2} u^*_{n+1} + bu_{n+1} u^*_{n-1} \\
&+ \frac{c}{\lambda} u_{n-1}u_{n-2})  + f_n \, ,
\end{aligned}
\label{eq:sabra}
\end{equation}
where $^*$ denotes the complex conjugate. 

The first term in the rhs of (\ref{eq:sabra}) is the dissipation with constant
viscosity $\nu$.  Notice that this term explicitly breaks the time
reversibility, i.e. the symmetry under the transformation $t\to -t$
and $u_n\to -u_n$, of the equation, as it does in the NSE.

The second, non-linear term, preserving the time-reversal symmetry,
couples velocity variables at different shells and is built in analogy
with the non-linear term of the NSE in Fourier space.  The
coupling is restricted to neighboring shells, owing to
the predominant locality of the energy cascade \cite{rose1978fully}.
Choosing the coefficients with the prescription
$a+b-c=0$ (in our simulations $a=1$ and $b=-1/2=-c$), the nonlinear
term preserves two quadratic invariants, i.e. energy $E=\sum_n |u_n|^2$
and helicity $H=\sum_n (-1)^n k_n |u_n|^2$, similarly to the NSE.

\begin{figure*}[t!]
\centering
\includegraphics[width=0.85\linewidth]{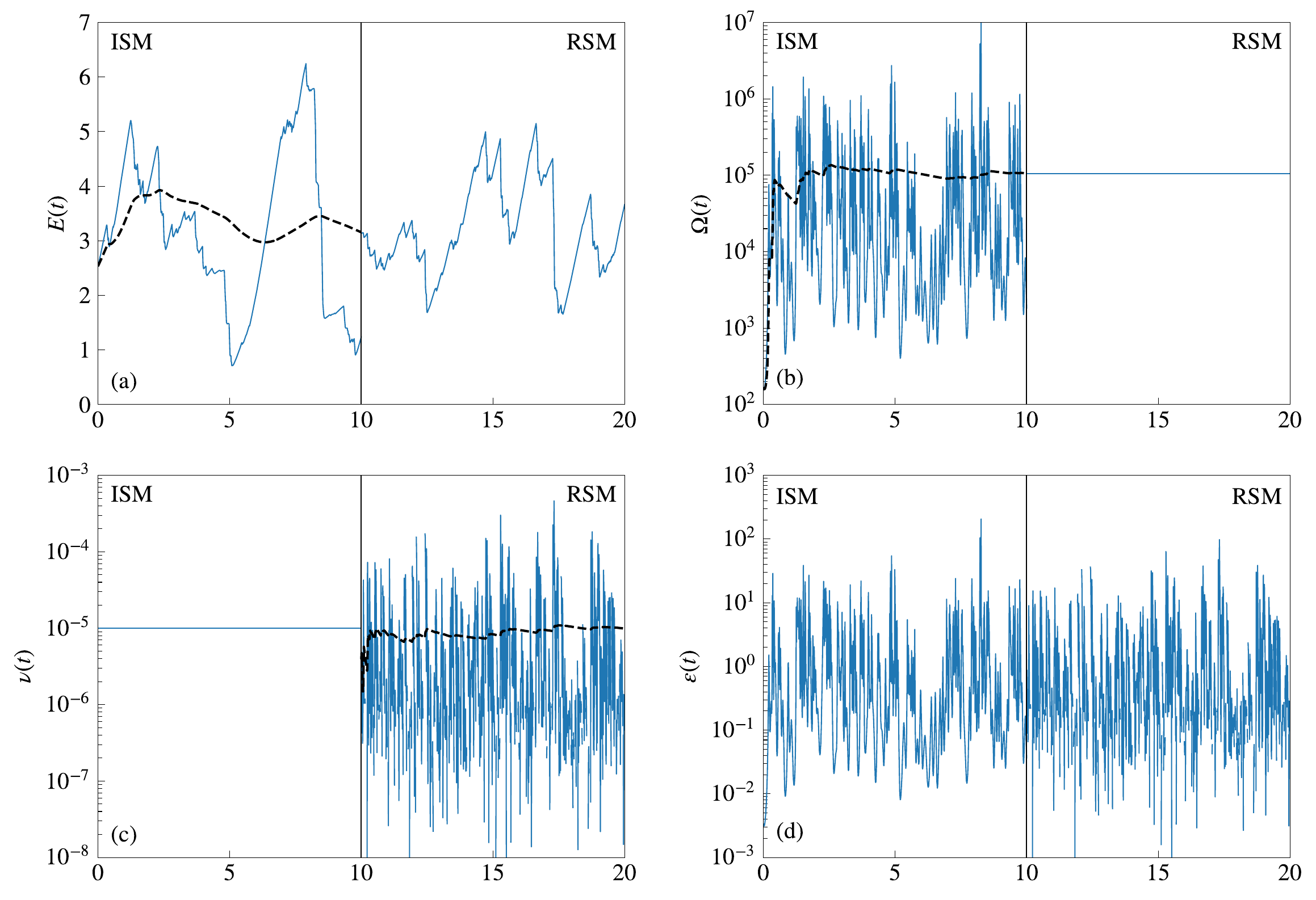}
\caption{Temporal dynamics of different observables measured during
  typical runs of both the ISM (\ref{eq:sabra}) (left side of each
  panel) and the RSM with viscosity given by (\ref{eq:nu_reversible})
  (right side of each panel). On the x--axis of all panels time is
  measured in simulation units. Panels: (a) total energy $E$; (b)
  total enstrophy $\Omega$; (c) viscosity coefficient $\nu$; (d)
  energy dissipation rate $\varepsilon(t) = 2 \nu \Omega$. Continuous
  lines represent instantaneous values, dashed lines represent running
  averages (in time). For details on simulations, see Appendix
  \ref{sec:app_procedure} and \ref{sec:app_params}.}
\label{fig:overview}
\end{figure*}

Finally, the last term $f_n$ represents the forcing, which injects
energy at an average rate $\langle\varepsilon\rangle=\langle\sum_n
\mathcal{R}\{f_n u_n^*\}\rangle$, where $\mathcal{R}$ denotes the real
part. In our simulations we considered a constant forcing, which
preserves the time-reversal symmetry, acting only on the large scales
(small wavenumbers) $f_n=f\delta_{n,0}$ with $f=const$.

\subsection{Reversible shell model (RSM)\label{sec:reversible}}
As discussed above the term $-\nu k_n^2 u_n$ in
Eq.~(\ref{eq:sabra}) explicitly breaks the time reversal symmetry. In
this section, we show how it can be modified by allowing the viscosity
to vary depending on the velocity variables in such a way that the
dynamics is (formally) time-reversible. In this way we can directly
probe the irreversibility due the non-equilibrium energy cascade.

The first proposal to modify the Navier-Stokes equation in such a way
to have a reversible dynamics is due to She and Jackson
\cite{she1993constrained} who introduced the constrained Euler
equation, in order to devise a new Large Eddy Simulation (LES) scheme,
by imposing a global constraint on the energy spectrum. On a more
theoretical ground, Gallavotti
\cite{gallavotti1996equivalence,Gallavotti1997Dynamical} proposed to
modify the dissipative term by letting the viscosity depend on the
velocity field in such a way to conserve a global quantity, e.g.
energy or enstrophy. The value of these quantities is then determined by the initial conditions which should be taken so that the total energy or enstrophy, depending on the chosen constraint, equal the average value obtained from a long integration of the irreversible model dynamics.
Gallavotti conjectured that these (formally)
reversible equations should be ``equivalent'', in the spirit of
equivalence of ensembles in equilibrium statistical mechanics, to the
(irreversible) NSE, at least in the limit of very high Reynolds
number. This idea was then tested, for some aspects, in 2D NSE
\cite{gallavotti2004lyapunov} and, more recently, in the Lorenz 1996
model \cite{gallavotti2014equivalence}, which can be thought as a
single scale shell model.

Here we apply these ideas to the shell model (\ref{eq:sabra}).  Past attempts to modify
(\ref{eq:sabra}) imposing the energy conservation have encountered
some difficulties in reproducing the dynamics of the original shell
model \cite{Biferale1998}. When fixing the energy, we found similar
difficulties. Briefly, the main problem is that, in the regime of
energy cascade, the value of the mean energy is essentially determined
at the integral (forcing) scales and is basically independent of the
viscosity.  Therefore, fixing the energy alone does not fix the
extension of the inertial range (\textit{viz.}  the Reynolds number).
On the other hand, fixing the enstrophy 
\begin{equation}
\Omega=\sum_n k_n^2 |u_n|^2 \, 
\label{eq:omega_def}
\end{equation}
enforces a constraint on the small scales so that once its value is
imposed via the initial condition the extension of the inertial range,
and thus the Reynolds number, results well defined also in the
reversible model.  By using (\ref{eq:sabra}) in the request
\begin{equation}
\dot{\Omega}=0 \, ,
\label{eq:omega_constant}
\end{equation}
one obtains the dynamical evolution for the time-reversible viscosity
\begin{equation}
\begin{aligned}
\nu_R(t)  &= \frac{\sum_n k_n^2 \mathcal{R}\{f_n u_n^*\}}{\sum_n k_n^4
  |u_n|^2} +\\ 
&+\frac{\sum_n\! k_n^3
  \!\left[a\lambda
    C_{3,n\!+\!1}\!+\!bC_{3,n}\!-\!\frac{c}{\lambda}C_{3,n\!-\!1}\right]}{\sum_n k_n^4
  |u_n|^2} \, ,
\end{aligned}
\label{eq:nu_reversible}
\end{equation}
where $C_{3,n}= -\mathcal{I}\{u_{n+1}u^*_nu^*_{n-1}\}$, and
$\mathcal{I}$ stands for the imaginary part.  It is worth noticing
that there are two terms in the rhs of Eq.~(\ref{eq:nu_reversible})
because enstrophy is both injected by the forcing (first term) and
produced by the nonlinear dynamics (second term).  Most importantly,
since $\nu_R$ is odd in the velocity variables, the modified
dissipative term $-\nu_R k_n^2 u_n$ preserves time reversal symmetry,
i.e. does not change sign for $t\to -t$ and $u_n\to -u_n$.

Being $\nu_R$ a variable quantity, the initial condition for the $u_n$
becomes the only way of controlling the separation between the
injection and dissipation scales in the system. Increasing the
enstrophy of the initial condition increases the separation of scales
and vice-versa. Further details on the simulation procedure can be
found in the Appendices \ref{sec:app_procedure} and
\ref{sec:app_params}.

We conclude the presentation of the reversible shell model by showing,
in Fig.~\ref{fig:overview}, the time evolution of some global
observables such as energy, enstrophy, energy dissipation and the
viscosity itself measured both in the ISM and RSM. As one can see, in
spite of the drastic change in the enstrophy and viscosity
(Fig. \ref{fig:overview}b,c) the qualitative features of energy and
energy dissipation are similar.  The highly intermittent behavior of
the energy dissipation, $\varepsilon(t)$, is qualitatively preserved
in RSM.  Notice that in the ISM the time-dependent energy dissipation
reads $\varepsilon(t)=\nu \Omega(t)$ while in the RSM it takes the form
$\varepsilon(t)=\nu_R(t)\Omega$, i.e. the quantity dependent on time
is enstrophy in the former and the viscosity in the latter with the
enstrophy $\Omega$ fixed at the average value obtained from the
ISM. Also, the time average of the variable viscosity
(\ref{eq:nu_reversible}) is approximately equal to the value of $\nu$
in the corresponding irreversible simulation, which is
  a prerequisite to have the dynamical equivalence between the two
  dynamics \cite{gallavotti1996equivalence}.

\begin{figure*}[h!]
\centering
\includegraphics[width=0.85\linewidth]{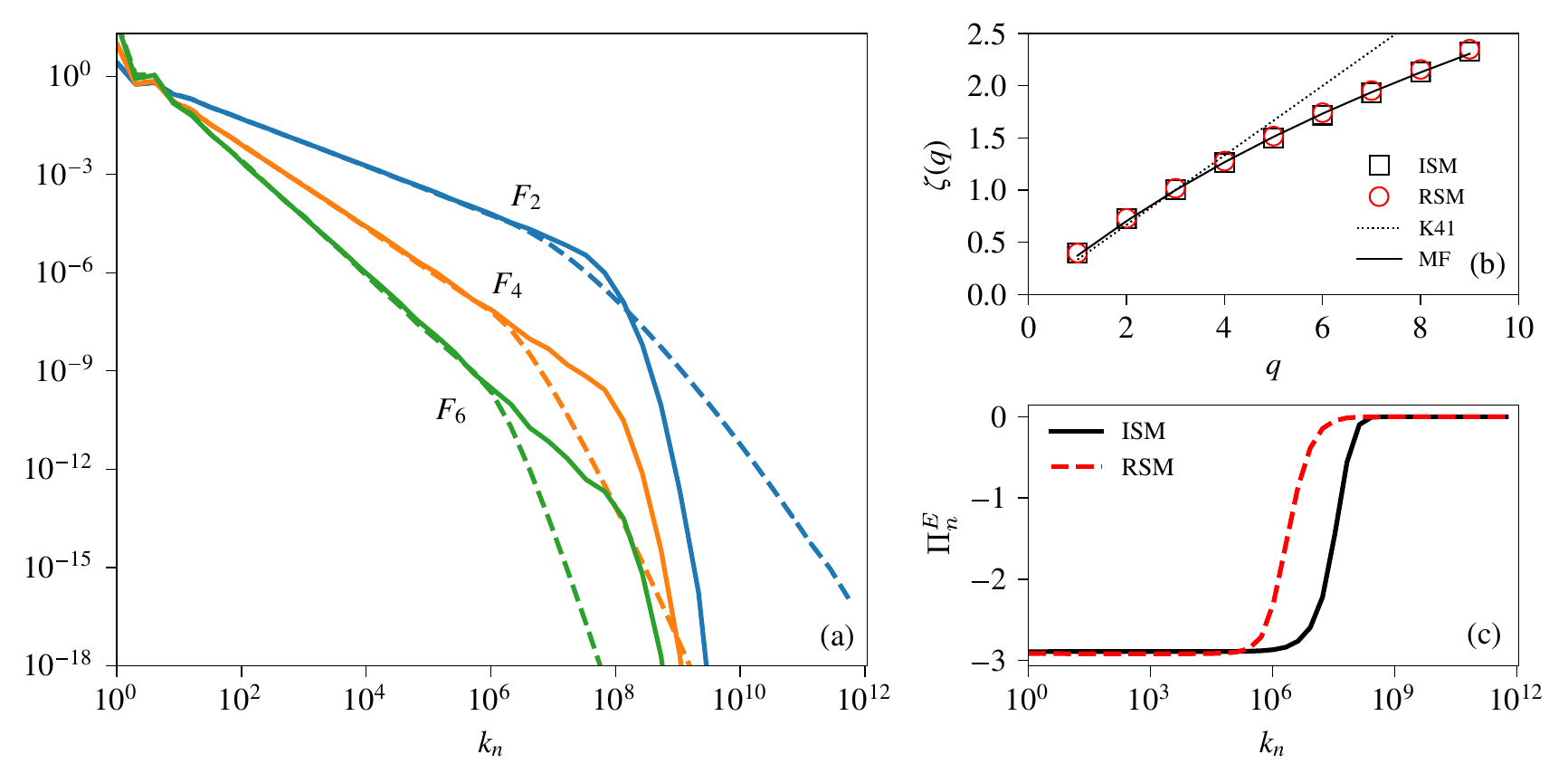}
\caption{Comparison between the two models. (a) Structure function
  $F_q(k_n)$ of order $q=2,4,6$ (as labeled) vs $k_n$, for the ISM
  (solid curves) and RSM (dashed curves). (b) Scaling exponents
  $\zeta(q)$ obtained by fitting the structure functions in the
  inertial range in the two models, compared with the K41 dimensional
  prediction ($q/3$) and the multifractal one (\ref{eq:zetaEMF}), see
  legend. (c) Energy flux $\Pi^E_n$ as a function of the scale in both
  models. In all panels the error bars are smaller than the symbols.
  For details on the parameters of simulations see Appendix
  \ref{sec:app_params} (parameter sets \textbf{I1} and \textbf{R1}).
\label{fig:SFrev}}
\end{figure*}
\subsection{Lagrangian statistics in shell models}
For shell models, lacking a spatial structure, there is not an obvious
recipe for introducing a Lagrangian velocity. However, as
observed in \cite{boffetta2002lagrangian}, the quantity
\begin{equation}
v(t) = \sum_n \mathcal{R}\{u_n(t)\}
\label{eq:vlag}
\end{equation}
can be regarded as a sort of Lagrangian velocity. The choice of the
real part is arbitrary, working with the imaginary part gives
equivalent results.

The rationale for (\ref{eq:vlag}) is that the Lagrangian velocity is
the superimposition of eddies at all scales, $u_n$ in the shell
models. Since the shell model is not affected by sweeping
\cite{bohr2005}, such a superposition is expected to reproduce the
statistics of velocity along the particle path.  Indeed it has been shown
that $v(t)$ as defined above shares many qualitative and quantitative
features of the Lagrangian velocity statistics of real 3D turbulent
flows \cite{boffetta2002lagrangian}. In particular, Lagrangian
structure functions have been shown to display a scaling behavior
with exponents deviating from the dimensional
prediction and quantitatively close to those observed in experiments
and simulations of the NSE
\cite{mordant2001measurement,chevillard2003lagrangian,biferale2008lagrangian,arneodo2008universal}.
Using (\ref{eq:vlag}) as a definition of Lagrangian velocity in
the shell models, we define the Lagrangian acceleration as
\begin{equation}
a(t) = \dot{v} = \sum_n \mathcal{R}\{\dot{u}_n(t)\}\,,
\label{eq:a}
\end{equation}
and the Lagrangian power
\begin{equation}
  p = va= \sum_n\mathcal{R}\{u_n\} \sum_m \mathcal{R}\{\dot{u}_m\}\,,
  \label{eq:p}
\end{equation}
whose statistics can be studied in oder to explore the issue of 
Lagrangian time irreversibility.

We notice that the constant forcing on the first shell, which is used
in our simulations, imposes a strong constraint on the phases of the
first shells, leading to $\langle v(t) \rangle \neq 0$.  Since, in
principle, this may induce some spurious effects on the asymmetry of
the power statistics, we have also tested our results with a
(time-reversible) stochastic forcing for which the statistics of $v$
is symmetric around $\langle v(t) \rangle = 0$, though
non-Gaussian. Since the results we present are independent of the
forcing choice, in the following we shall only show the constant
forcing results, for a comparison with the other choice the reader may
consult \cite{cencini2017time}.

\section{Energy cascade and anomalous scaling in the reversible shell model \label{sec:confronto}}
The modified viscosity (\ref{eq:nu_reversible}) can be interpreted
within the framework of large eddy simulations as an effective model
for small scale dissipation. In this respect it is worth mentioning
that also for the NSE several time-reversible LES model have been
proposed \cite{carati2001modelling,fang2012time,Jimenez2015}.  It is
thus important to verify whether and to what extent the RSM is able to
reproduce the inertial range physics of the ISM.  In particular, here,
we study the scaling behavior of velocity structure functions that for
shell models read
\cite{jensen1991intermittency,biferale2003shell,bohr2005}
\begin{equation}
F_q(k_n)=\langle|u_n|^q\rangle\sim k_n^{-\zeta(q)}\,.
\label{eq:sfshell}
\end{equation}
and the energy spectrum defined as $E_n \equiv F_2(k_n)=\langle |u_n|^2 \rangle$.
For the standard Sabra shell model it has been shown
\cite{Lvov_1998_improved_shellmodels} that the exponents $\zeta(q)$
deviate from the dimensional (Kolmogorov 1941, K41) prediction, i.e.
$\zeta(q)\neq q/3$, and are quantitatively close to the exponents
of the Eulerian structure functions observed in experiments and
simulations of the NSE.

In Fig.~\ref{fig:SFrev}a, we compare the structure functions
$F_q(k_n)$ for $q=2,4$ and $6$ obtained from both ISM and RSM. As one
can see their inertial-range scaling behavior is essentially
indistinguishable. This is further confirmed in Fig.~\ref{fig:SFrev}b
where we compare the scaling exponents of the structure functions
$\zeta(q)$ obtained by fitting the structure functions in both
models. In Fig.~\ref{fig:SFrev}b we also show that the scaling
exponents are very well described by the multifractal formula (see
Appendix~\ref{app:MF} for a brief summary of the MF model for turbulence)
\begin{equation}
\zeta(q)= \inf_{h} \{hq+3-D(h)\}\,,
  \label{eq:zetaEMF}
\end{equation}
where for $D(h)$ we used a log-Poisson model [see Eq.~(\ref{dofh})].

The constancy of the energy
flux, $\Pi_n^E$, through the scale $k_n$, displayed in
Fig.~\ref{fig:SFrev}c, confirms that in both models a direct energy
cascade is taking place. We remark, however, that the reversible shell
model displays a slightly reduced inertial range, as inertial scaling
disappears a few shells before its irreversible
counterpart. 
A major difference between the two models is
apparent in the dissipative range. Indeed the RSM shows a non trivial
behavior at the scales where the ISM is exponentially damped by the
fixed-viscosity dissipation. 
\subsection{Small scale behavior of the RSM}

The reversible and irreversible shell models display different statistics 
at small scales, due to the different dissipative schemes.
In the following we focus on this range of scales by
looking at the energy and enstrophy spectra at varying the Reynolds
number, i.e. the extension of the inertial range.

In the ISM, we observe an exponential suppression of turbulent
fluctuations after the inertial range of scales, i.e. above the
Kolmorogov wavenumber, $k_\eta \approx
(\nu^3/\langle\varepsilon\rangle)^{-1/4}$. Conversely, in the RSM, we
can distinguish an additional range of scales for $k>k_\eta$
characterized by a scaling close to a power law as clear from
Fig.~\ref{fig:Confronto_spettri_rev}a, where we show the energy
spectrum at increasing $\Omega$.  At even larger wavenumbers, this
power-law decay is followed by an exponential suppression, which is
not visible in Fig.~\ref{fig:Confronto_spettri_rev}a due to the
limited resolution but is clearly observed in simulations at smaller
$\Omega$ (not shown).

As shown in Fig.~\ref{fig:Confronto_spettri_rev}b, the post-inertial
range of scales shows a trend toward constancy of enstrophy at
different $k_n$, suggesting equipartition of enstrophy and $E_n \sim
k_n^{-2}$.  Simulations at higher resolution (high $N$) and high
values of $\Omega$ are computationally very demanding, due to the
stiffness of ODE (\ref{eq:sabra}) and its numerical instability, so we
were not able to explore higher values of $\Omega$ and determine
unambiguously whether an effective equipartition of enstrophy is
reached in the limit $\Omega \rightarrow \infty$. A further
complication in understanding the physics of this range of scales is
that both enstrophy equipartition and enstrophy cascade (constant
flux) are characterized by the same energy spectrum scaling $E_n \sim
k_n^{-2}$, making it difficult to predict the physical mechanism behind
the observed dynamics from only looking at the spectrum. To
disentangle the two possibilities, one would have to look at the
enstrophy flux, however, at difference with $E$ or $H$, the enstrophy
$\Omega$ is not an invariant for the non-linear term of equation
(\ref{eq:sabra}), and its time-derivative cumulated on the first $M$
shells cannot be interpreted as a rate of transfer. In our simulations
we found that the enstrophy dynamics is dominated by the balancing
between the enstrophy generated by the non-linear interactions and the
enstrophy dissipation.

\begin{figure}
\centering
\includegraphics[width=0.99\linewidth]{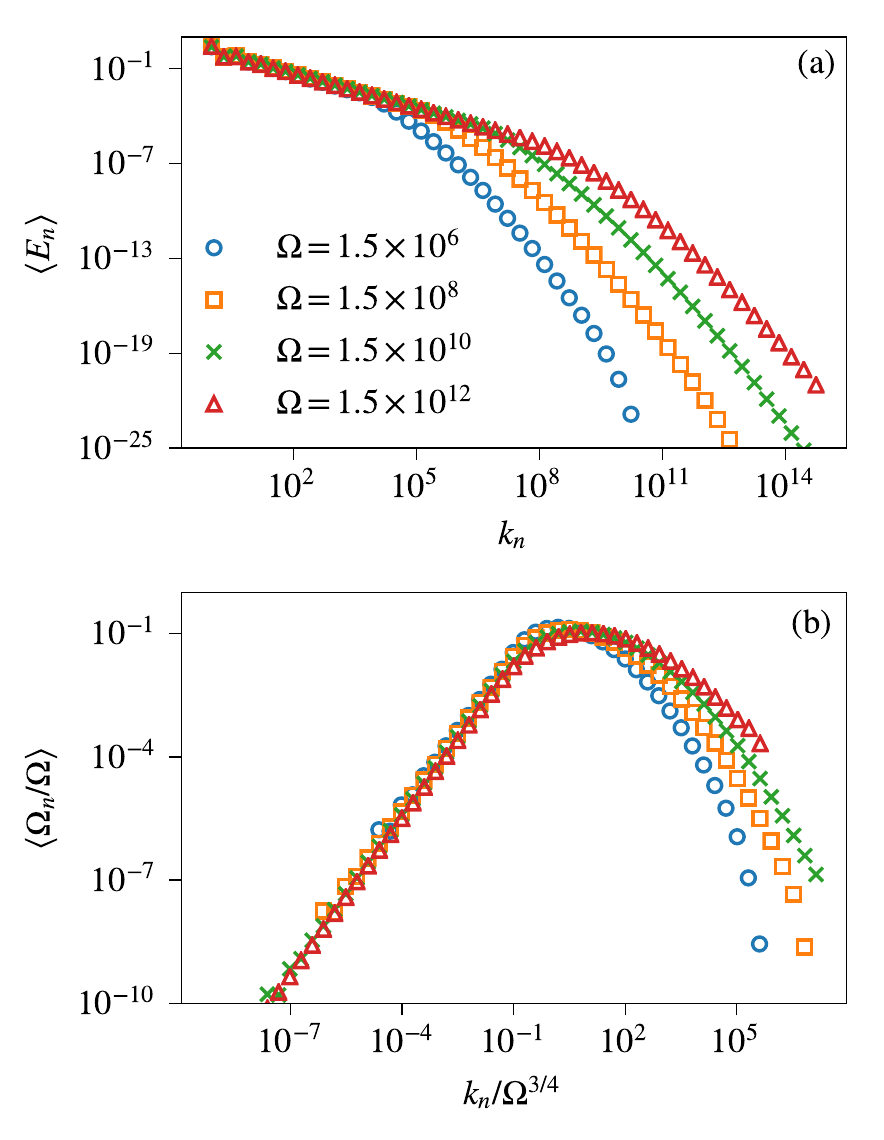}
\caption{Energy (a) and enstrophy spectra (b) of the RSM at varying
  the total enstrophy $\Omega$. The enstrophy spectra in (b) have been
  rescaled in order to keep the Kolmogorov length-scale $O(1)$.
  Notice that in the RSM the Kolmogorov scale can be defined as
  $k_\eta\approx (\langle \nu \rangle ^3 / \langle \varepsilon
  \rangle)^{-1/4}\sim \langle \varepsilon \rangle^{-1/2}
  \Omega^{3/4}$, where we used that $\langle \nu \rangle=\langle
  \varepsilon\rangle / \Omega$.  Errors, not shown, are of the same
  order of the symbol size or less. For details on simulations, see
  Appendix \ref{sec:app_a} (parameter sets \textbf{R2-5}).}
\label{fig:Confronto_spettri_rev}
\end{figure}

Regardless of the underlying physical mechanism, the existence of a
post-inertial range of scales suggests that the energy dissipation
statistics of the RSM could be substantially different from that of
the ISM.  We thus studied the moments of the energy dissipation at
varying the Reynolds number in both models. More specifically, we
studied how the moments depend on the Taylor scale Reynolds number
defined as $Re_\lambda=E/\sqrt{\nu\langle\varepsilon\rangle}$, i.e. as
the ratio between the large scale time scale,
$T_L=E/\langle\varepsilon\rangle$, and the small scale Kolmogorov time
scale, $\tau_\eta=\sqrt{\nu/\langle\varepsilon\rangle}$.  For the ISM,
the moments of the energy dissipation are known to follow a power-law
scaling on $Re_\lambda$ \cite{boffetta2000energy}
\begin{equation}
\label{eq:momeps}
\langle \varepsilon^q\rangle \sim
Re_\lambda^{\chi(q)}
\end{equation}
with the exponents $\chi(q)$ in agreement with the multifractal model as (see also Appendix~\ref{app:MF})
\begin{equation}
\chi(q)=\sup_h\left\{2\frac{D(h)-3-(3h-1)q}{1+h}\right \}\,,
\label{eq:MT_resulteps}
\end{equation}
where $D(h)$ is the same function used for the structure functions (\ref{eq:zetaEMF}).

Since, in the RSM, the viscosity (\ref{eq:nu_reversible}) can assume
negative values, we studied the moments of the absolute value of the
energy dissipation $\langle |\varepsilon|^q \rangle$ (we also checked
that moments preserving the sign, such as $\langle
|\varepsilon|^{q-1}\varepsilon\rangle$ give the same results, not
shown).  In Fig.~\ref{fig:epsilonmom} we show the exponents obtained
by fitting the scaling behavior (\ref{eq:momeps}) for the moments of
energy dissipation for both the RSM and ISM, together with the
prediction (\ref{eq:MT_resulteps}). As one can see in the RSM the
moments are definitely different from the ISM values, which are well
predicted by (\ref{eq:MT_resulteps}). In particular, the exponents of
the RSM are smaller, meaning that the intermittency of $\varepsilon$
is weaker in the reversible model.

\begin{figure}
\centering
\includegraphics[width=0.99\linewidth]{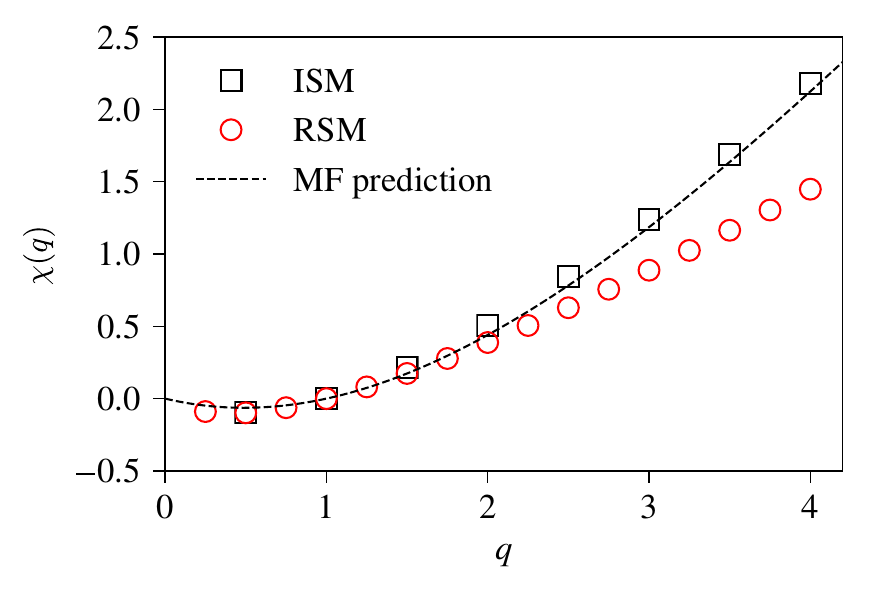}
\caption{Scaling exponents of the moments of the energy dissipation
  (\ref{eq:momeps}) for the RSM (circles), and for the ISM
  (squares). The dashed line represents the MF prediction
  (\ref{eq:MT_resulteps}). Errors, not shown, are of the same order of
  the symbol size or less. For details on simulations, see Appendix
  \ref{sec:app_a} (parameter sets \textbf{I2-9} and \textbf{R6-14}).}
\label{fig:epsilonmom}
\end{figure}

\section{Lagrangian Power statistics and time irreversibility\label{sec:lagpow}}
It is useful to start this Section by briefly summarizing previous
findings on Lagrangian power statistics in turbulence. As mentioned in
the introduction, by inspecting both experimental and numerical
trajectories of Lagrangian tracers Xu et al. \cite{xu2014} discovered
that time increments of Lagrangian kinetic energy are negatively
skewed and that this skewness persists for the time derivatives,
i.e. for the Lagrangian power (\ref{eq:pnse}).
Such skewness is
directly linked to the time irreversibility of the tracer dynamics, as
it means that the probability of gaining and losing kinetic energy is not the same, though
$\langle p\rangle=0$ (by stationarity).  In particular,
they found that approximately:
\begin{equation}
\langle p^2 \rangle \simeq \langle\varepsilon\rangle^2 Re_\lambda^{4/3}\,, \quad
\langle p^3 \rangle \simeq -\langle\varepsilon\rangle^3 Re_\lambda^{2}\,.
\label{eq:mom2e3xu}
\end{equation}
The above results convey two messages. First, the probability density
function  of $p$ is skewed, with $\langle p^3\rangle /\langle
p^2\rangle ^{3/2}\approx const<0$ suggesting that time-irreversibility
is robust and persists in the limit $Re_\lambda \to
\infty$. Second, the exponents $4/3$ and $2$, which approximately
describe the scaling behavior of the second and third moment, strongly
deviate from the dimensional prediction based on K41
theory, according to which
\begin{equation}
\langle p^q\rangle/\langle\varepsilon\rangle^q \propto Re_\lambda^{q/2} \, , 
\label{eq:K41_lagpower}
\end{equation}
meaning that the Lagrangian power is strongly intermittent.

It has been shown, in \cite{cencini2017time}, that the deviations from (\ref{eq:K41_lagpower}) can be understood within the framework of the multifractal
model for turbulence (see also Appendix~\ref{app:MF}).  
In particular,  the MF model predicts that
\begin{equation}
\langle p^q \rangle\! \sim\! \langle\varepsilon\rangle^q Re_\lambda^{\alpha(q)}
\label{eq:MF_lagpower}
\end{equation}
 with
\begin{equation}
  \alpha(q)=\sup_h\left\{2\frac{(1-2h)q-3+D(h)}{1+h}\right\}\,.
\label{eq:MF_lagpower_exponent}
\end{equation}
In \cite{cencini2017time} it is also shown that, defining
the Lagrangian power as in (\ref{eq:p}), the (irreversible) shell
model displays an intermittent statistics for $p$, but at variance with 
NS-turbulent data, deviations from the  prediction  (\ref{eq:MF_lagpower_exponent}) are
present, at least in the statistical asymmetries of the power pdf. In
this section, we broaden the investigation comparing Lagrangian power statistics in both the ISM and RSM.

\subsection{Moments and asymmetry of Lagrangian power}
For both the ISM and RSM the Lagrangian
power is defined according to Eq.~(\ref{eq:p}).
\begin{figure}
\centering
\includegraphics[width=0.99\linewidth]{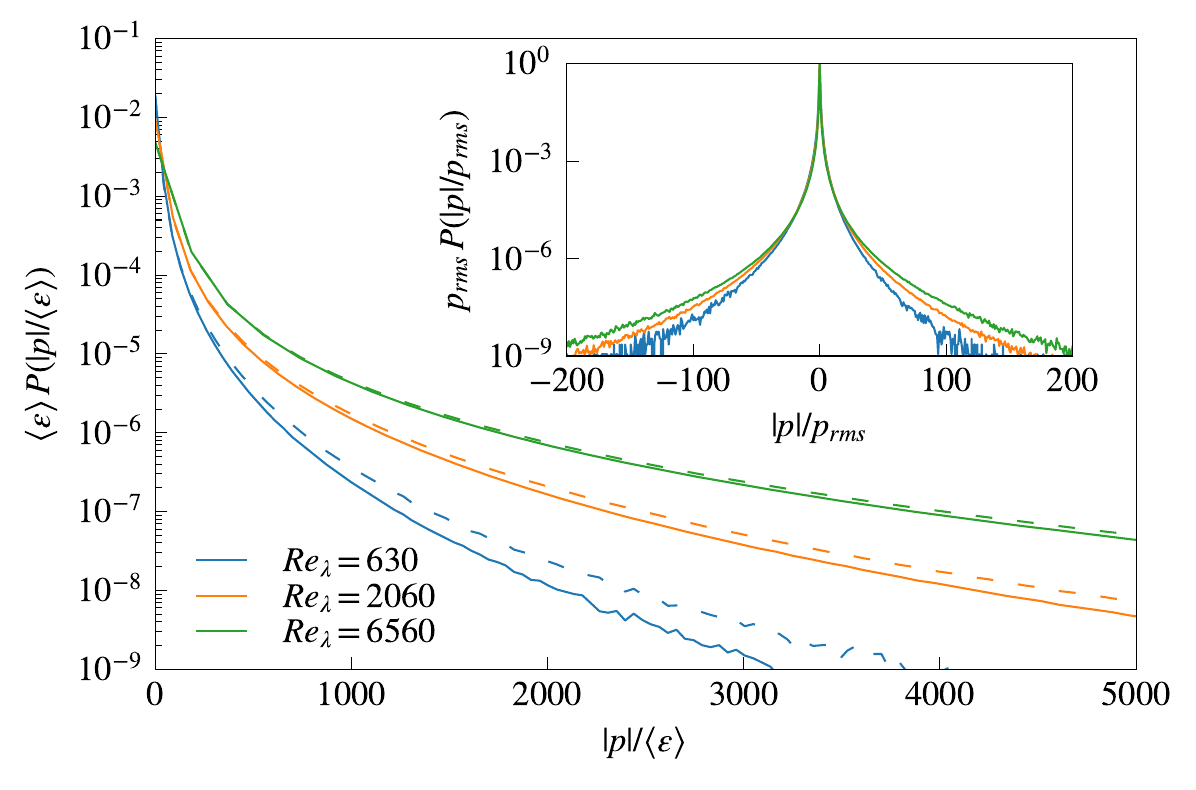}
\caption{Probability density function of the Lagrangian power
  normalized by the average energy input rate, $p/\langle
  \varepsilon\rangle$, at three values of $Re_\lambda$ for the ISM. To highlight
  tail asymmetries, the pdf is plot against $|p|/\langle
  \varepsilon\rangle$, the positive/negative tail is in solid/dashed
  lines. Inset: the three pdfs of the main plot normalized with
  $p_{rms}=\langle p^2\rangle^{1/2}$, the curves do not overlap which
  is the signature of intermittency in the power statistics. For
  details on simulations, see Appendix \ref{sec:app_a} (parameter sets
  \textbf{I4}, \textbf{I6}, \textbf{I8}).}
\label{fig:pdfP}
\end{figure}
As discussed above, time irreversibility reveals itself in the odd order
moments of the power that are sensitive to the asymmetries in the
tails of the pdf of power.  Such asymmetries are shown in
Fig.~\ref{fig:pdfP} for different values of $Re_\lambda$. The absence
of collapse onto a unique curve for the pdf of $p/\langle
p^2\rangle^{1/2}$ (shown in the inset) highlights the presence of
intermittency in the statistics of $p$.  Here, following
\cite{cencini2017time}, in order to probe the scaling behavior of the
symmetric and asymmetric component of the statistics we introduce two
non-dimensional moments:
\begin{equation}
S_q=\frac{\langle
|p|^q\rangle}{\langle\varepsilon\rangle^q}; \quad A_q=\frac{\langle
p|p|^{q-1}\rangle}{\langle\varepsilon\rangle^q}\,.
\label{eq:defmom}
\end{equation}
Clearly the latter vanishes for a symmetric (time-reversible) pdf.  

\begin{figure*}
\centering
\includegraphics[width=0.85\linewidth]{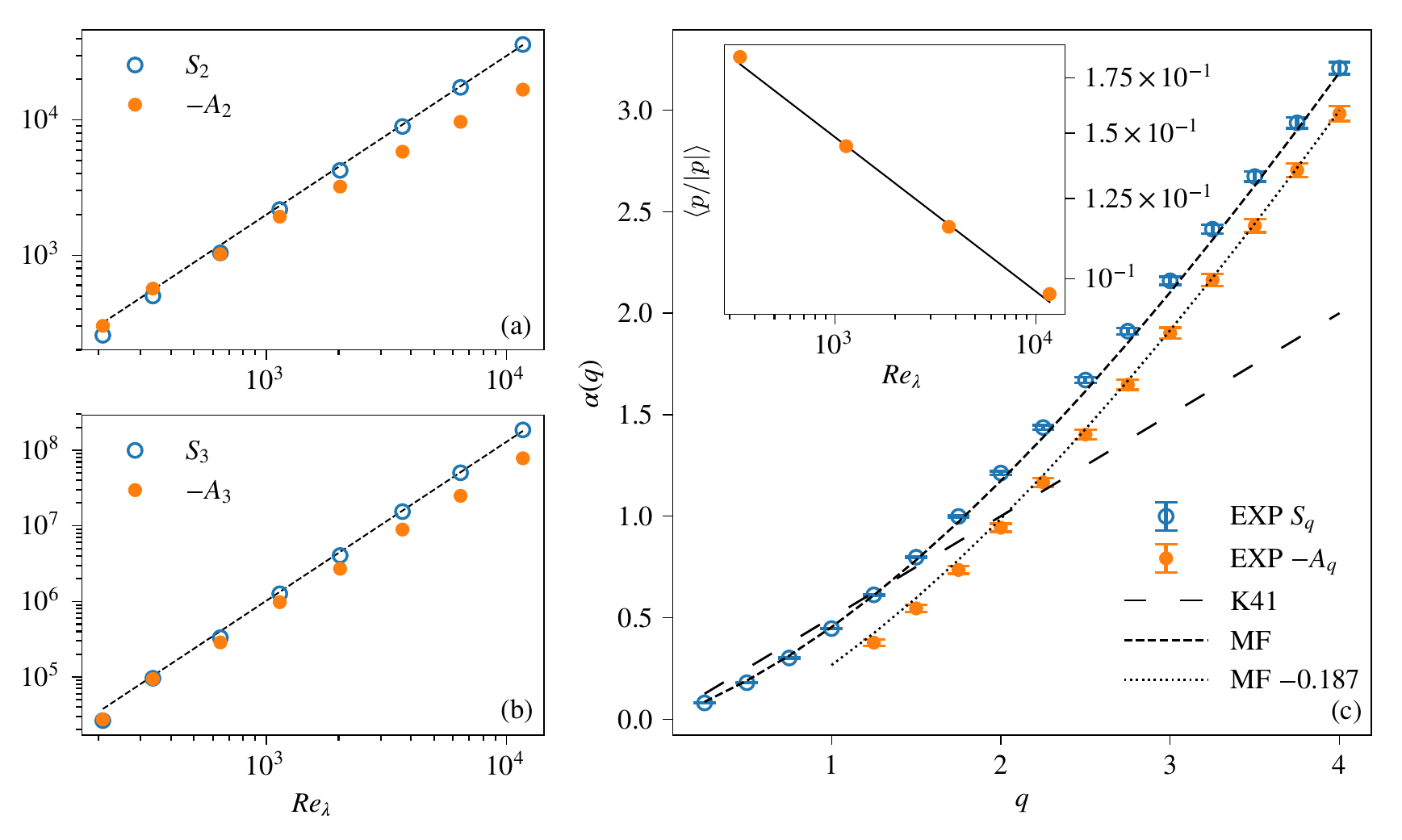}
\caption{Lagrangian power statistics in the ISM. Power moments $S_q$
  and $-A_q$ (see legend) as a function of $Re_\lambda$ for (a) $q=2$
  and (b) $q=3$. The curves for $-A_q$ have been shifted vertically to
  highlight the difference with respect to $S_q$.  The black solid
  line shows the MF prediction
  (\ref{eq:MF_lagpower})--(\ref{eq:MF_lagpower_exponent}). Panel (c):
  exponents for the $Re_\lambda$ dependence fitted from $S_q$ and
  $-A_q$ compared with K41 (\ref{eq:K41_lagpower}) and MF predictions
  (\ref{eq:MF_lagpower})--(\ref{eq:MF_lagpower_exponent}). Inset:
  $Re_\lambda$-dependence of $\langle p/|p|\rangle\propto
  Re_\lambda^{-\mu}$ with $\mu\approx 0.187(7)$ as obtained by a best
  fit shown as a black line.  Where error bars are not shown, it means
  that they are smaller or equal to the symbol size. For details on
  simulations, see Appendix \ref{sec:app_a} (parameter sets
  \textbf{I2-9}).
\label{fig:SMirrev}}
\end{figure*}
\begin{figure*}
\centering \includegraphics[width=0.85\linewidth]{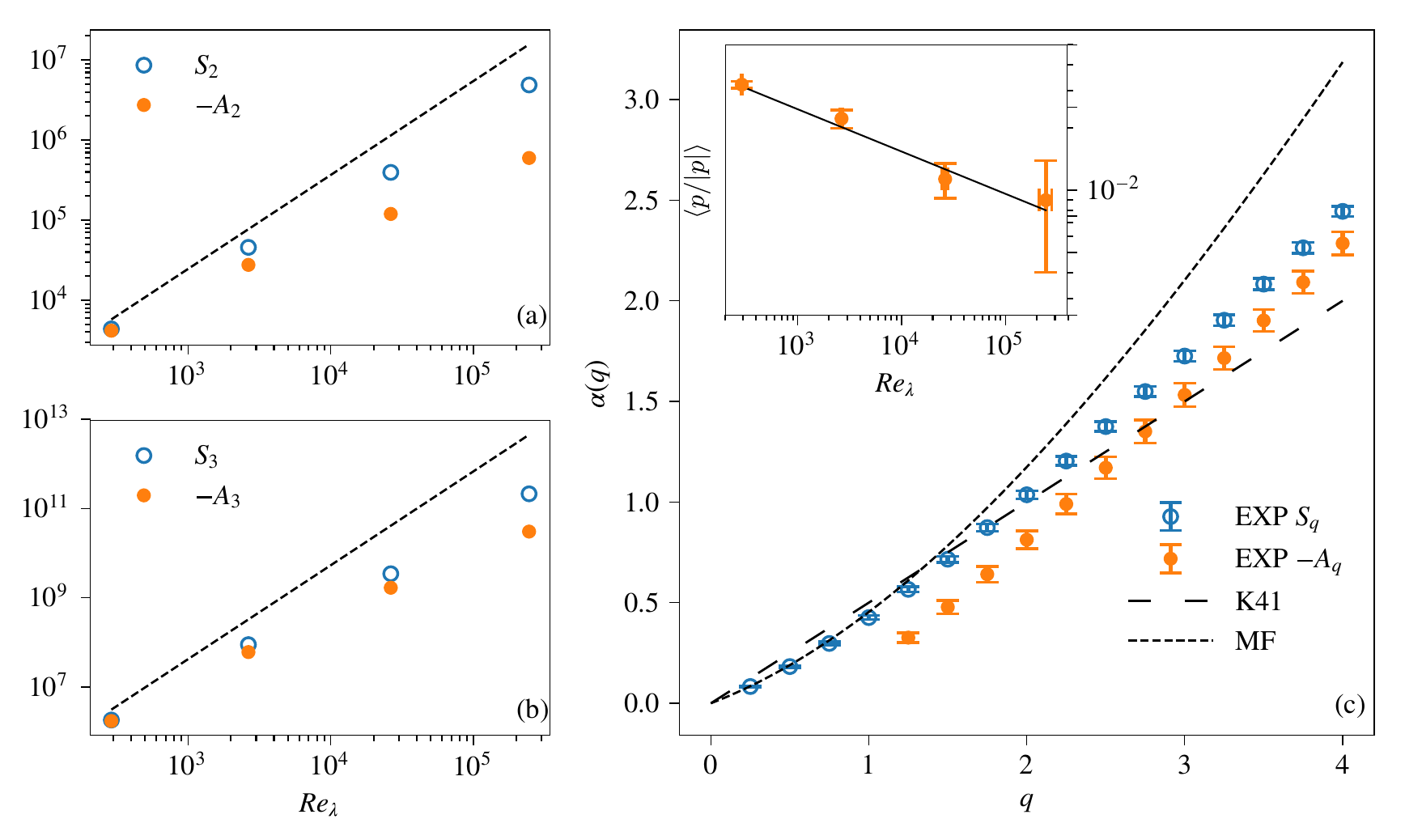}
\caption{Lagrangian power statistics in the RSM. Panels (a), (b) and
  (c) present the same quantities as in Fig.~\ref{fig:SMirrev} but for
  the RSM.  Inset of panel (c): $Re_\lambda$-dependence of $\langle
  p/|p|\rangle\propto Re_\lambda^{-\mu}$ with $\mu\approx 0.20(7)$ as
  obtained by a best fit shown as a black line.  For details on
  simulations, see Appendix \ref{sec:app_a} (parameter sets
  \textbf{R2-5}).}
\label{fig:SMrev}
\end{figure*}

The main results on the Lagrangian power moments are summarized in
Fig.~\ref{fig:SMirrev} and Fig.~\ref{fig:SMrev} for the ISM and RSM,
respectively. In Fig.~\ref{fig:SMirrev}a,b (Fig.~\ref{fig:SMrev}a,b)
we show the second and third moments for the ISM (RSM),
respectively. Two observations are in order.
\begin{enumerate}
\item As for the ISM, the symmetric moments $S_q$ are in excellent
  agreement with the scaling behavior in $Re_\lambda$ predicted by the
  MF model obtained using (\ref{eq:MF_lagpower_exponent}) with the
  $D(h)$ given by (\ref{dofh}) (see Fig.~\ref{fig:SFrev}b).
  Conversely, deviations from the MF prediction are evident in the
  RSM.
\item For both models, the asymmetric moments $A_q$ are negative
  (positive) for $q>1$ ($q<1$) (we recall that $A_1=0$ by
  stationarity).  The non-vanishing values of $A_q$ for $q\neq 0$ are
  the signature of time reversal symmetry breaking.  In both models,
  the scaling behavior of $A_q$ is definitively different from that of
  $S_q$. In particular, the exponents are smaller and thus the
  asymmetry in the tails appears to be subleading with respect to the
  symmetric component.
\end{enumerate}
The second observation implies that the generalized skewnesses
$\tilde{S}_q=-A_q/S_q$, which measures the scaling ratio between the
asymmetric and the symmetric components of the statistics at varying
the order, are decreasing functions of $Re_\lambda$. This suggests
that there is a statistical recovery of the time reversal symmetry in
the limit of infinite $Re_\lambda$, at variance with what observed in
NS turbulence \cite{xu2014,xu2014b,cencini2017time}. It is important
to stress that the decay of the generalized skewness does not imply
the decay of standard measures of skewness \cite{cencini2017time},
such as e.g. $\langle p^3\rangle /\langle p^2\rangle^{1/2}$, which may
be still growing with $Re_\lambda$ due to intermittency corrections,
(see \cite{BV2001} for a similar issue in the problem of statistical
recovery of isotropy).

Figure~\ref{fig:SMirrev}c (Fig.~\ref{fig:SMrev}c) summarizes the
results concerning the scaling exponents of the moments of power in
the ISM (RSM). We can see the excellent agreement between the fitted
exponents for $S_q$ of the ISM and the MF prediction. Strong
deviations from the MF prediction are evident for the RSM, which is
characterized by exponents smaller than predicted, denoting a less
intermittent statistics. This behavior is consistent with the
observation made for the energy dissipation
(Fig.~\ref{fig:epsilonmom}).  This points to a major role played by
the contribution of the dissipative terms to the Lagrangian power of
shell models. To verify this, for the ISM, we decomposed the power in
its contributions due to forcing, $p_f=v \sum_n \mathcal{R}\{f_n\}$,
dissipation, $p_d=-v \nu \sum_n k_n^2 \mathcal{R}\{u_n\}$, and
nonlinear terms, $p_{nl}=v \sum \mathcal{R} \{ i k_n (a\lambda u_{n+2}
u^*_{n+1} + bu_{n+1} u^*_{n-1}) \}$, where $p=p_f+p_{d}+p_{nl}$. We
found that $\langle p_d^2\rangle/\langle p^2\rangle\approx 1$ and
$\langle p_{nl}^2\rangle/\langle p^2\rangle\approx 2$ independently of
$Re_\lambda$, which confirms that the dissipative and non-linear
contributions scale as the total power and that they are of the same
order. This is at odds with what has been observed in DNS of turbulent
flows \cite{xu2014b}, where the statistics is dominated by the
pressure gradients, i.e. by the nonlinear terms, and the dissipative
contribution was found to be subleading in terms of scaling and less
intense with respect to the nonlinear one.

Figures~\ref{fig:SMirrev}c and \ref{fig:SMrev}c also show the
exponents obtained by fitting the scaling behavior of the
antisymmetric moments. For both ISM and RSM these
exponents can be linked to the symmetric exponents by a rigid
shift, i.e. 
\begin{equation}
-A_q \sim S_q Re_\lambda^{-\mu}\,.
\label{eq:shift}
\end{equation}
We found this relation
to be consistent with the assumption that, in terms of scaling behavior, 
there is a decoupling between the absolute value of the power and its
sign, i.e. $A_q \sim \langle p/|p|\rangle S_q$. Indeed for both
models, as shown in the insets of Figs.~\ref{fig:SMirrev}c and \ref{fig:SMrev}c, we found 
\begin{equation}
\langle p/|p|\rangle \sim
Re_\lambda^{-\mu}\,,
\label{eq:sign}
\end{equation}
with $\mu \approx 0.18$ ($0.2$) for the ISM (RSM).  At present, this
is just an observation of which we do not have a clear understanding.
It should be remarked that the relation of (\ref{eq:shift}) with
(\ref{eq:sign}) shows that the multifractal model is not completely
failing in reproducing the asymmetries of the power statistics, and
that the scaling behavior of the asymmetries is compatible, modulo the
cancellation exponent $\mu$ (see also \cite{ott1992sign}), with the
multifractal phenomenology. We emphasize that in DNS of the
Navier-Stokes equations \cite{cencini2017time} there is no evidence of
a cancellation exponent different from zero, suggesting that the
asymmetry persists also in the infinite Reynolds number limit.

\section{Discussions and Conclusions \label{sec:conclusions}}
In this paper we have introduced a time-reversible shell model for
turbulence, obtained by modifying the dissipative term of the
so-called Sabra model, allowing the viscosity to vary in such a way as
to maintain the total enstrophy constant.  In spite of the formal time
reversibility of the model we found that the dynamics spontaneously
breaks the time reversal symmetry selecting an attractor onto which
irreversibility manifests in the asymmetry of the Lagrangian power
statistics.

A detailed quantitative comparison between the reversible and
irreversible (original) shell models has shown that the dynamics of
the former well reproduce the inertial range physics of the latter,
indeed the structure functions of the two models are indistinguishable
in the inertial range. On the contrary, the modified viscous term of
the reversible model is responsible for important modifications in the
physics below the Kolmogorov scale. While the irreversible model at
these scales is characterized by an energy spectrum with an
exponential fall-off, in the reversible model an intermediate
range characterized by a close-to equipartition of enstrophy physics
appears. The difference between the two models in this range of scale
is responsible for a different statistics of the energy
dissipation. As for the Lagrangian power statistics, we found that even
though qualitatively the two models display the same features,
quantitative details are different. In particular, the exponents
characterizing the scaling behavior of the moments of power of the
reversible model are smaller than those of the irreversible
model. These differences are consistent with those observed for the
energy dissipation and have, possibly, a similar origin in the non
trivial physics of the reversible model below the Kolmogorov scale.

As for the irreversible shell model, consistently with our previous
observations \cite{cencini2017time}, we found that independently of the
nature of the forcing, the (time-reversible) symmetric statistics of
the power statistics are well captured by the multifractal model while
deviations are present for the (time-irreversible) asymmetric
component, which is characterized by smaller exponents. However,
numerical evidence suggests that these deviations can be traced back
to the Reynolds dependence of the sign of the power (cancellation
exponent \cite{ott1992sign}). This indicates that the bulk part of the
statistics is well captured by the multifractal model. Time-reversible
sub-grid models for Large Eddy Simulations of the NSE might be
important to better capture backscatter events where the energy is
locally transferred from small to large scales in
turbulence, i.e. when an inverse energy transfer is observed. The
issue is particularly subtle considering that there is not a unique
meaning of local energy transfer in the configuration space and that
some of the inverse transfer events are probably simply due to large
instantaneous fluctuations disconnected from any robust transfer
mechanism \cite{chaodyn}.

\begin{acknowledgements}
We thank R. Benzi, M. Sbragaglia and G. Gallavotti for fruitful
discussions.  We acknowledge support from the COST Action MP1305
``Flowing Matter''.  LB and MDP acknowledge funding from ERC under the
EU $7^{th}$ framework Programme, ERC Grant Agreement No 339032.
\end{acknowledgements}

\section*{Authors contribution statement}
All the authors conceived the study. MDP performed the
simulations of RSM, MC performed simulations of the ISM. All the
authors analyzed the data, discussed the results and wrote the
manuscript.

\appendix
\section{Details on simulations of RSM}
\label{sec:app_a}
\subsection{General procedure for reversible simulations}
\label{sec:app_procedure}

In our simulations we always started by integrating equations
(\ref{eq:sabra}) with constant viscosity over a time period long
enough to guarantee stationarity of the dynamics and the convergence
of the time averages of all the quantities measured.

Denoting with  $\langle E_n \rangle |_\nu$ the average energy associated to
the $n$-th shell of an irreversible shell-model simulation with
constant viscosity $\nu$, we define the initial condition for the
corresponding reversible simulation as
\begin{equation}
\label{eq:ic_rev}
u_n(t=0)|_\Omega = [\langle E_n \rangle |_\nu]^{1/2} \, [\cos(\zeta_n) + i \sin(\zeta_n)] \, ,
\end{equation}
where the $\zeta_n$ are random angles. This definition guarantees that
the reversible simulation will start with a total energy $E$ equal to
the average energy of the corresponding irreversible run, and an
enstrophy $\Omega$ (conserved in this case) equal to the average
enstrophy of the corresponding irreversible run. For the irreversible
simulations, the initial velocity field is chosen with random values
on the first $6$ shells, and an initial energy $E(t=0) \sim 1$.

In all cases the time-integration algorithm we used is a modified
fourth order Runge-Kutta scheme with explicit integration of the
viscous term. 

We averaged our measurements on an ensemble of $\sim 10$
simulations, differing in the choice of the initial conditions, for
both the irreversible and the reversible models. In the reversible
case, it is possible to build ensembles of simulations characterized
by the same ``Reynolds number'' by picking different values for the
$\zeta_i$ in (\ref{eq:ic_rev}), being the separation of scales
effectively controlled by the ratio $\Omega / E$.

\subsection{Parameters used}
\label{sec:app_params}

Here we report the sets of parameters used for the simulations presented in this paper. The \textbf{I} sets are for simulations of the ISM, the \textbf{R} sets are for simulations of the RSM. $N$ is the number of shells; $|f|$ is the magnitude of the forcing; $\nu$ is the  value of the constant viscosity; $\Omega$ is the value of enstrophy; $dt$ is the integration timestep used; $T$ is the total time of integration, summed over all the simulations in the ensemble. The average energy and the big eddy turnover time are always $O(1)$.\smallskip

\begin{center}
\begin{tabular}{ c c c c c c }
set          & $N$  & $|f|$  & $\nu$                 & $dt$                 & $T$  \\ 
\hline
\textbf{I1}  & $40$ & $1$    & $10^{-10}$            & $2 \times 10^{-8}$   & $\sim 3000$  \\ 
\textbf{I2}  & $30$ & $\sqrt{2}$ & $10^{-4}$             & $10^{-4}$            & $\sim 10^{6}$\\
\textbf{I3}  & $30$ & $\sqrt{2}$ & $3.16 \times 10^{-5}$ & $10^{-4}$            & $\sim 10^{6}$\\
\textbf{I4}  & $30$ & $\sqrt{2}$ & $10^{-5}$             & $5 \times 10^{-5}$   & $\sim 10^{6}$\\
\textbf{I5}  & $30$ & $\sqrt{2}$ & $3.16 \times 10^{-6}$ & $2.5 \times 10^{-5}$ & $\sim 10^{6}$\\
\textbf{I6}  & $30$ & $\sqrt{2}$ & $10^{-6}$             & $1.5 \times 10^{-5}$ & $\sim 10^{6}$\\
\textbf{I7}  & $30$ & $\sqrt{2}$ & $3.16 \times 10^{-7}$ & $10^{-5}$            & $\sim 10^{6}$\\
\textbf{I8}  & $30$ & $\sqrt{2}$ & $10^{-7}$             & $5 \times 10^{-6}$   & $\sim 10^{6}$\\
\textbf{I9}  & $30$ & $\sqrt{2}$ & $3.16 \times 10^{-8}$ & $2.5 \times 10^{-6}$ & $\sim 10^{6}$\\
\hline
\end{tabular}

\medskip

\begin{tabular}{ c c c c c c }
set          & $N$  & $|f|$ & $\Omega$              & $dt$               & $T$  \\ 
\hline
\textbf{R1}  & $40$ & $1$   & $1.44 \times 10^{10}$ & $5 \times 10^{-9}$ & $\sim 720$   \\
\textbf{R2}  & $35$ & $1$   & $1.45 \times 10^{6}$  & $5 \times 10^{-7}$ & $\sim 12000$ \\
\textbf{R3}  & $45$ & $1$   & $1.44 \times 10^{8}$  & $5 \times 10^{-8}$ & $\sim 4500$  \\
\textbf{R4}  & $50$ & $1$   & $1.44 \times 10^{10}$ & $10^{-8}$          & $\sim 720$   \\
\textbf{R5}  & $50$ & $1$   & $1.46 \times 10^{12}$ & $10^{-9}$          & $\sim 405$   \\
\textbf{R6}  & $25$ & $1$   & $1.29 \times 10^{4}$  & $2 \times 10^{-6}$ & $\sim 130000$\\
\textbf{R7}  & $25$ & $1$   & $4.38 \times 10^{4}$  & $2 \times 10^{-6}$ & $\sim 120000$\\
\textbf{R8}  & $25$ & $1$   & $1.41 \times 10^{5}$  & $2 \times 10^{-6}$ & $\sim 190000$\\
\textbf{R9}  & $35$ & $1$   & $4.29 \times 10^{5}$  & $5 \times 10^{-7}$ & $\sim 19000$ \\
\textbf{R10} & $35$ & $1$   & $1.46 \times 10^{6}$  & $5 \times 10^{-7}$ & $\sim 10000$ \\
\textbf{R11} & $35$ & $1$   & $4.26 \times 10^{6}$  & $5 \times 10^{-7}$ & $\sim 10000$ \\
\textbf{R12} & $35$ & $1$   & $1.46 \times 10^{7}$  & $5 \times 10^{-7}$ & $\sim 8000$  \\
\textbf{R13} & $45$ & $1$   & $4.30 \times 10^{7}$  & $5 \times 10^{-8}$ & $\sim 9000$  \\
\textbf{R14} & $45$ & $1$   & $1.46 \times 10^{8}$  & $5 \times 10^{-8}$ & $\sim 3800$  \\
\hline
\end{tabular}
\end{center}

\section{Multifractal model for Eulerian and Lagrangian statistics\label{app:MF}}

Here we briefly recall the basic ideas on the 
\textit{multifractal model} (MF) of turbulence
\cite{FP1985,benzi1984multifractal,frish_turbulence} for the Eulerian statistics.

According to the MF model, Eulerian velocity
increments at inertial scales are characterized by a local H\"older
exponent $h$, i.e. $\delta_r u \sim U_L(r/L)^h$ ($L$ and $U_L$
denoting the large scale and the associated velocity, respectively),
whose probability ${\cal P}(h) \sim r^{3-D(h)}$ depends on the fractal
dimension $D(h)$ of the set where $h$ is observed. Thus the Eulerian
structure functions can be written as
$$
\langle (\delta_r u)^q\rangle \sim U_L^q \int_{h\in
  \mathcal{I}} dh \left(\frac{r}{L}\right)^{hq+3-D(h)} \sim U_L^q
\left(\frac{r}{L}\right)^{\zeta(q)}\,,$$
where a saddle point approximation for $r\ll L$ gives
\begin{equation}
  \zeta(q)= \inf_{h\in \mathcal{I}} \{hq+3-D(h)\}\,.
  \label{eq:zetaEMFapp}
\end{equation}
As for the Eulerian $D(h)$, in this paper we assume a Log-Poisson 
functional form \cite{she1994universal}
\begin{equation} 
D(h) = \frac{3 (h-h_0)}{ \log(\beta)} \left[\log \left( \frac{3
    (h_0-h)}{d_0 \log(\beta)}\right) -1\right] +3 -d_0\,.
    \label{dofh}
\end{equation}
with $h_0=1/9$ and $d_0 = (1-3h_0)/(1-\beta)$ is fixed by imposing the
exact relation $\zeta(3)=1$. For the shell model considered in this paper
choosing $\beta=0.6$ one obtains an excellent fit for the scaling
exponents of the structure functions (see Fig.~\ref{fig:SFrev}).

In the MF framework, it is also possible to derive a prediction for 
the moments of the energy dissipation $\langle \varepsilon^q\rangle$
We can indeed estimate the (fluctuating) energy dissipation as
$\varepsilon=(\delta_\eta u)^3 /\eta$. According to the MF model 
$\eta=(\nu L^h/U_L)^{1/(1+h)}$, and thus we have
\begin{equation}
\varepsilon=U_L^{4/(1+h)} \nu^{(3h-1)/(1+h)}L^{-4h/(1+h)}\,.
\label{eq:epsMF}
\end{equation}
Then using the same procedure that led to (\ref{eq:zetaEMF}) one can
derive the scaling behavior of $\langle \varepsilon^q\rangle$ as a
function of $Re_\lambda \sim \nu^{-1/2}$, namely
\begin{equation}
\strut{\hspace{-0.2cm}}\langle \varepsilon^q\rangle \sim
Re_\lambda^{\chi(q)} \; \mathrm{with} \; \chi(q)=\sup_{h\in \mathcal{I}}\left\{2\frac{D(h)-3-(3h-1)q}{1+h}\right \}\,.
\label{eq:resulteps}
\end{equation}
\textit{En passant} notice that the $4/5$ law, namely the fact that
$\zeta_3=1$ implies $\chi(1)=0$, i.e. the dissipative anomaly.

The multifractal model has been
successfully extended to describe also several aspects of Lagrangian
statistics, such as velocity
\cite{borgas1993,boffetta2002lagrangian,chevillard2003lagrangian,arneodo2008universal},
acceleration \cite{borgas1993,biferale2004} and Lagrangian power
\cite{cencini2017time}.  Here we briefly recall the main steps.

To connect temporal velocity differences $\delta_\tau v$ over a time
lag $\tau$, along fluid particle trajectories to equal time spatial
velocity differences $\delta_r u$, in
Refs.\cite{borgas1993,boffetta2002lagrangian} it was noticed that
$\delta_\tau v$ should receive the main contribution from eddies at a
scale $r$ such that $\tau \sim r/\delta_r u$.  This implies
$\delta_\tau v \sim \delta_r u$ that establishes the bridge between
Lagrangian and Eulerian quantities linking times and length scales:
\begin{equation}
\tau \sim T_L ({r}/{L})^{1-h}
\label{eq:ELbridge}
\end{equation}
where $T_L=L/U_L$ is the eddy turnover time of the large scales.  The
bridging relation (\ref{eq:ELbridge}) provides a way to derive the
expression for the scaling exponents of the Lagrangian structure
function $S_q^L(\tau)=\langle(\delta_\tau v)^q\rangle\sim U_L^q
(\tau/T_L)^{\zeta_L(q)}$ with 
\begin{equation}
\xi(q)=\inf_{h\in\mathcal{I}}\{[hq+3-D(h)]/(1-h)\}\,,
\end{equation}
which with the same $D(h)$ used for the Eulerian statistics provides
exponents in agreement with the shell model
\cite{boffetta2002lagrangian} and with experimental and DNS data of
NS-turbulence \cite{chevillard2003,biferale2004,arneodo2008universal}.
In the same spirit, MF can be used to describe acceleration statistics
noticing that $a \sim {\delta_{\tau_\eta} v}/{\tau_\eta}$, which
yields \cite{biferale2004}
\begin{equation}
  a \sim \nu^{(2h-1)/(1+h)} U_L^{3/(1+h)} L^{-3h/(1+h)}\,.
  \label{eq:acc2}
\end{equation}
The above expression, together with the definition of Lagrangian
power, $p=av$ and recalling that $\nu \sim Re_\lambda^{-1/2}$, can be
used to predict how the power moments $\langle p^q\rangle$ depend on
$Re_\lambda$, which is \cite{cencini2017time}
\begin{equation}
\langle p^q \rangle\! \sim\! \langle \varepsilon\rangle^q Re_\lambda^{\alpha(q)}
\end{equation}
 with
\begin{equation}
  \alpha(q)=\sup_{h\in \mathcal{I}}\left\{2\frac{(1-2h)q-3+D(h)}{1+h}\right\}\,.
\label{eq:expo}
\end{equation}
It is worth remarking that the scaling in $Re_\lambda$ is essentially
carried by the acceleration, meaning that $\langle a^q\rangle \sim
Re_\lambda^{\alpha(q)}$. In other terms a by product of the above
analysis is that we should expect $\langle p^q \rangle \sim \langle
a^q\rangle $ as far as scaling behavior is concerned.

We conclude noticing that, though asymmetries can be in principle
introduced in the MF formalism (see, e.g.,
\cite{chevillard2012phenomenological}), the above derivations bear no
information of statistical asymmetries in the statistics, therefore
all the predictions should be understood as holding for the asymmetric
components of the statistics, i.e. for the moments of the absolute
values of the relevant quantities.


\end{document}